\documentclass[fleqn,10pt]{olplainarticle}
\title{Modular and Automated Workflow for Streamlined Raman Signal Analysis}
\author[1]{Mykyta Kizilov}
\author[1]{Vsevolod Cheburkanov}
\author[2]{Joseph Harrington}
\author[1, 2*]{Vladislav V. Yakovlev}
\affil[1]{Department of Biomedical Engineering, Texas A\&M University, College Station, TX 77843, USA}
\affil[2]{Department of Physics and Astronomy, Texas A\&M University, College Station, TX 77843, USA}
\affil[*]{Email: yakovlev@tamu.edu}

\keywords{Raman spectroscopy, automated preprocessing, Voigt peak fitting, baseline correction, noise suppression}

\begin{abstract}
Raman spectroscopy is a powerful tool for material characterization. However, careful preprocessing is required for the identification and handling of noise, baseline drift, and random spikes. This paper presents a comprehensive approach to generating and preprocessing Raman spectra. Additionally, we describe methods for fitting Voigt peaks to the spectrum to determine peak parameters. The effectiveness of these methods is demonstrated using both synthetic and real Raman spectra, with code provided in an open-source GitHub repository.
\end{abstract}
\begin{document}
\flushbottom
\maketitle
\thispagestyle{empty}

\section{Introduction}

Raman spectroscopy is a highly sensitive, non‑destructive technique for probing molecular vibrations, rotations, and other low‑frequency modes in materials \cite{opilik2013modern}. By detecting the inelastic scattering of monochromatic light, it allows the detailed characterization of molecular compositions, structures, along with interactions, thereby offering insights into molecular conformations and bonding environments \cite{long2002raman}. This particular technique continues to be widely adopted in numerous scientific fields such as chemistry \cite{esmonde2017raman}, biology \cite{krafft2006biomedical}, physics, and materials science \cite{weber2013raman}, due to its ability to capture the molecular fingerprints of various samples.

One of the most important applications of Raman spectroscopy is in protein analysis. It can provide crucial information about amino acid composition and secondary structural properties with excellent precision \cite{altangerel2024novel}. The ability to estimate protein structure and dynamics without needing extensive sample preparation makes Raman spectroscopy especially helpful for studying biological macromolecules. The technique is sensitive to small molecular changes, such as conformational or frequency shifts or variations in the Raman bands. These features help with the determination of protein structure, the orientation of side chains, and the observation of side chain interactions, contributing to a much deeper understanding of protein functionality \cite{kengne2012protein}.

Nevertheless, despite its numerous advantages, Raman spectroscopy data are subjected to several artifacts and noise sources that prevent accurate information extraction. Common problems include baseline drift due to sample fluorescence  \cite{lieber2003automated}, random cosmic ray interference, and instrumental noise. These challenges are especially apparent when analyzing complicated biological samples, where small spectral shifts are crucial for correct interpreting  \cite{gebrekidan2016shifted, esmonde2017raman}. Therefore, effective preprocessing techniques, such as baseline correction, noise suppression, and peak fitting, are essential to enhance the quality and reliability of the analysis \cite{moreira2008raman, mykraman}.

Conventional preprocessing methods depend on manual tuning of parameters which is both time-consuming and inconsistent across datasets. Furthermore, modeling Raman spectral peaks may be complicated by the convolution of instrumental and natural broadening effects, which calls for advanced peak fitting approaches, like the Voigt profile, to precisely capture the peak shapes \cite{ida2000extended}. These limitations highlight the need for robust and automated preprocessing workflows that can handle common artifacts with minimal manual intervention.

Analyzing data is another significant challenge in Raman Spectroscopy applications. The untargeted nature of Raman data demands sophisticated computational methods \cite{ryabchykov2019analyzing} to interpret complex spectral profiles. Conventional chemometric techniques, like principal component analysis (PCA) and partial least squares regression  \cite{dunn1989principal}, are widely used for spectral analysis. But recent trends emphasize the use of machine learning (ML) and deep learning (DL) models to enhance data modeling, classification, and anomaly detection \cite{gil2023denoising, ralbovsky2020towards}. Such models can automate baseline correction, noise reduction, and peak fitting to reduce manual intervention and increase reproducibility.

While ML and DL approaches have potential, their practical implementation is still restricted by the need for large labeled datasets and computing resources \cite{qi2023recent}.  It's also hard to integrate these techniques in real time applications because most algorithms require computationally intensive post-processing steps. These gaps will call for initiatives from across spectroscopy, information science and engineering disciplines to produce faster, more effective workflows.

Regardless of the accessibility of different preprocessing methods, challenges remain in attaining high precision, particularly for spectra with complicated noise patterns and overlapping peaks. Recent developments have explored wavelet transforms \cite{chen2018adaptive} and adaptive filtering algorithms \cite{jordan1989adaptive} to address random noise in spectra, but their application is often limited by the need for careful parameter tuning. Furthermore, PCA methods have been applied to isolate noise components, but they can inadvertently eliminate subtle spectral features critical for biological analysis \cite{dunn1989principal}. More advanced denoising approaches using convolutional neural networks have shown potential \cite{gil2023denoising}, though they require large, labeled datasets to perform effectively. These issues underscore the trade-offs between simplicity, accuracy, and computational efficiency in spectral noise reduction workflows.

Baseline correction also presents significant challenges. Traditional polynomial and asymmetric least squares (ALS) methods are widely used to model fluorescence backgrounds, yet their performance can degrade with high-fluorescence samples or in spectra with broad peaks \cite{eilers2005baseline, ye2020baseline}. Additionally, spectral features close to the noise floor may be masked during baseline correction, leading to incomplete peak detection. Automated peak-fitting algorithms, such as those based on Lorentzian or Voigt profiles, are particularly prone to errors when dealing with heavily overlapping peaks or weak signals \cite{sundius1973computer}. As such, balancing peak resolution, computational speed, and the need for accurate detection remains a persistent challenge, particularly in real-time or high-throughput applications. This motivates the need for more robust and adaptive workflows that integrate multiple techniques to improve the quality and reliability of Raman spectral analysis.

\begin{figure}[h!]
\centering
\includegraphics[width=0.8\linewidth]{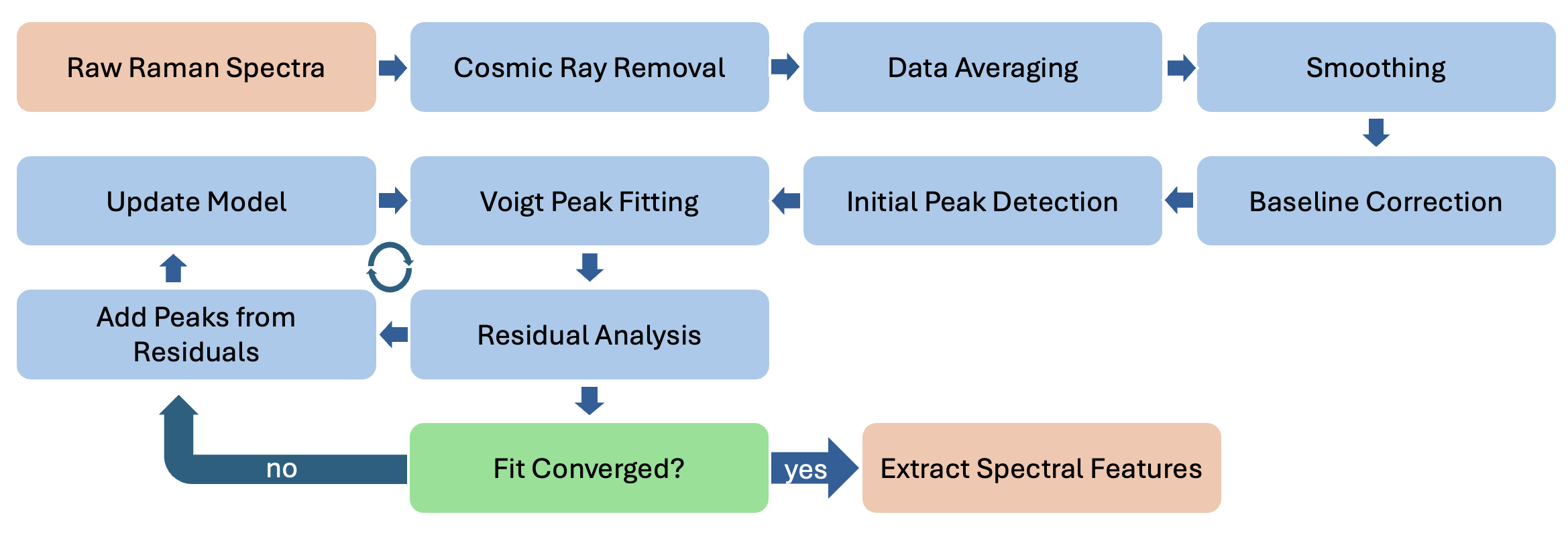}
\caption{Automated Raman analysis workflow: from raw spectra acquisition through cosmic ray removal, baseline correction, smoothing, peak detection, and Voigt fitting. The pipeline is modular and fully scriptable.}
\label{fig:raman_workflow}
\end{figure}

To address these issues, we present a systematic, automated workflow for Raman data preprocessing and analysis (Fig.~\ref{fig:raman_workflow}). This pipeline is easy to use, computationally efficient, and does not require large labeled datasets. Our goal is to develop a workflow that is simple to apply, consistent across datasets, and reproducible in different experimental contexts. The workflow efficiently suppresses noise, corrects baseline drift, and accurately fits spectral peaks—enhancing the precision of Raman spectral analysis. 

The proposed workflow is validated using both synthetic and experimental Raman spectra, demonstrating its effectiveness in mitigating common artifacts and extracting key spectral features. Furthermore, we provide open-source code through a GitHub repository to ensure reproducibility and to encourage further development in this area.

By prioritizing transparency, minimal parameter tuning, and modular design, this framework is intended as a general-purpose starting point for Raman analysis. It is readily extensible for more specialized tasks or integration with machine learning pipelines. Voigt functions are used to model peak shapes, ensuring reliable peak characterization. We validate the framework using both synthetic and experimental data and provide open-source code via GitHub to encourage reproducibility and further development.

\section{Synthetic Raman Data}

Generating synthetic Raman spectra is essential for testing and validation of data preprocessing and analysis techniques. This process entails the generation of spectra that accurately replicate real-world Raman signals, incorporating Voigt peaks, polynomial baseline components, and arbitrary noise and cosmic ray signals.

\subsection{Wavenumber Grid}
Raman spectra are typically represented in wavenumber coordinates, which ensures that the positions of spectral features are independent of the exact wavelength value of the excitation light in nanometers. For the purposes of our algorithm, an evenly spaced grid of wavenumber coordinates is selected and defined as:

\begin{equation}
\text{Wavenumbers} = \left\{ \nu_i \mid \nu_i = \nu_{\text{min}} + i \cdot \Delta\nu, \, i = 0, 1, \ldots, n-1 \right\}
\end{equation}

where \( \nu_{\text{min}} \) and \( \nu_{\text{max}} \) denote the minimum and maximum wavenumber values, respectively, establishing the boundaries of the desired spectral range.\\
\( n \) is the total number of points. The step size \(\Delta\nu\) is calculated as:

\begin{equation}
\Delta\nu = \frac{\nu_{\text{max}} - \nu_{\text{min}}}{n - 1}
\end{equation}

\subsection{Simulated Voigt Peaks}
In Raman spectroscopy, the Raman effect results from the inelastic scattering of photons by molecular vibrations. The energy difference between the incident and scattered photons corresponds to the vibrational energy levels of the molecules, producing characteristic peaks in the Raman spectrum. The observed peak shapes are influenced by both natural broadening, due to the finite lifetime of the excited vibrational states, and instrumental broadening, resulting from the spectrometer's resolution and other factors. The Voigt function, which is a convolution of a Lorentzian and a Gaussian function, is commonly used to model these peaks, as it accounts for both types of broadening.

The Lorentzian component represents the natural broadening, while the Gaussian component accounts for instrumental and Doppler broadening effects. Using the Voigt profile allows for a more accurate representation of the Raman peaks, especially in high-resolution spectra where both types of broadening are significant.

Spectral features of Raman spectra are modeled using a linear combination of Voigt functions shown here:

\begin{equation}
I_{\text{peaks}}(\nu) = \sum_{j=1}^{n_{\text{peaks}}} V(\nu, A_j, \nu_{0,j}, \sigma_j, \gamma_j)
\end{equation}

where \( V(\nu, A_j, \nu_{0,j}, \sigma_j, \gamma_j) \) is the Voigt function defined as the convolution of a Gaussian and a Lorentzian function:

\begin{equation}
V(\nu, A, \nu_0, \sigma, \gamma) = A \cdot \int_{-\infty}^{\infty} \frac{\gamma}{\pi} \frac{1}{(\nu' - \nu_0)^2 + \gamma^2} \cdot \frac{1}{\sigma \sqrt{2\pi}} e^{-\frac{(\nu - \nu')^2}{2\sigma^2}} d\nu'
\end{equation}

Alternatively, the Voigt function can be expressed using the real part of the Faddeeva function \( w(z) \):

\begin{equation}
V(\nu, A, \nu_0, \sigma, \gamma) = A \cdot \frac{\operatorname{Re}\left[ w\left( \frac{\nu - \nu_0 + i \gamma}{\sigma \sqrt{2}} \right) \right]}{\sigma \sqrt{2\pi}}
\end{equation}

In practice, numerical algorithms are used to compute the Voigt function efficiently \cite{letchworth2007rapid}.

\subsection{Polynomial Baseline}
The fundamental principles of inelastic light scattering indicate that a scattered photon can gain or lose energy during the scattering event, resulting in the generation of spectral components at shorter and longer wavelengths, known as Anti-Stokes and Stokes shifts, respectively. This report focuses primarily on the analysis of the Stokes component of Raman spectra, particularly in biological samples.

Biological samples often exhibit pronounced fluorescence when exposed to monochromatic light excitation. Given that fluorescence signals are typically red-shifted relative to the excitation wavelength and display broad spectral features, they can frequently obscure the Raman signals of interest. To address this issue, we simulate the broad fluorescence background using high-order polynomials.

In this model, the fluorescence baseline is described by a polynomial function of the wavenumber, expressed as:

\begin{equation}
B(\nu) = \sum_{k=0}^{d} a_k \nu^k
\end{equation}

where \( d \) is the degree of the polynomial and \( a_k \) are the coefficients, scaled appropriately to reflect realistic background levels.

\subsection{Noise and Spike Simulation}
In contemporary spectrometers, Raman spectra are detected using advanced cameras equipped with CCD or CMOS sensors. To accurately select an appropriate noise distribution for our model, it is crucial to understand the specific sources of noise associated with the detector in question.

There are several possible sources of noise in Raman data: cosmic ray artifacts, read noise, dark current noise, and shot noise.

In simulating Raman spectra, noise is typically modeled as Gaussian noise because it reflects the characteristics of the primary sources of random noise in the measurement process \cite{janesick1987scientific}.

Shot noise is described by a Poisson distribution. A Poisson distribution can be approximated by a Gaussian distribution if the mean photo-electron count detected by the camera pixel is sufficiently high. Consequently, when the spectral irradiance is high enough, the total noise in the measurement can be effectively modeled using a single additive Gaussian distribution \cite{barton2018algorithm}.\\

Gaussian noise is introduced to replicate the random fluctuations commonly observed in experimental data. This noise follows a normal distribution described here:

\begin{equation}
N(\nu) \sim \mathcal{N}(0, \sigma)
\end{equation}

where \( \sigma \) is the standard deviation that represents the noise level.

Cosmic ray signals in recorded spectra are observed as uncorrelated, high-amplitude spectral lines that often exhibit perceived spectral widths much narrower than those physically resolvable by the spectrometer. These signals result from high-energy particles that interact with the camera sensor and can occur at arbitrary points within the recorded spectrum at arbitrary times during data acquisition.

In the described methodology, cosmic ray signals are modeled as stochastic events with a specified probability $p$, and their amplitudes are drawn from a uniform distribution:

\begin{equation}
S(\nu) = \begin{cases}
M, & \text{with probability } p \\
0, & \text{otherwise}
\end{cases}
\end{equation}

where \( M \) is a random magnitude within a specified range. The magnitude of these spikes can vary widely. Some cosmic ray spikes may have amplitudes of only a few hundred counts, while others can reach several thousand counts. Also, different detectors may have varying sensitivities to cosmic rays. Because of that, there are no exact rules for choosing the magnitude range for \( M \). For simulation, it is recommended to choose a magnitude range similar to what you observe from data from an experimental Raman setup.

\subsection{Synthetic Spectrum Composition}
The final synthetic Raman spectrum is a sum of Voigt peaks, polynomial baseline, Gaussian noise, and random spikes:

\begin{equation}
I_{\text{total}}(\nu) = I_{\text{peaks}}(\nu) + B(\nu) + N(\nu) + S(\nu)
\end{equation}

This model creates spectra that have realistic features, making them suitable for testing various preprocessing and analysis methods.

\begin{figure}[h!]
\centering
\includegraphics[width=0.5\linewidth]{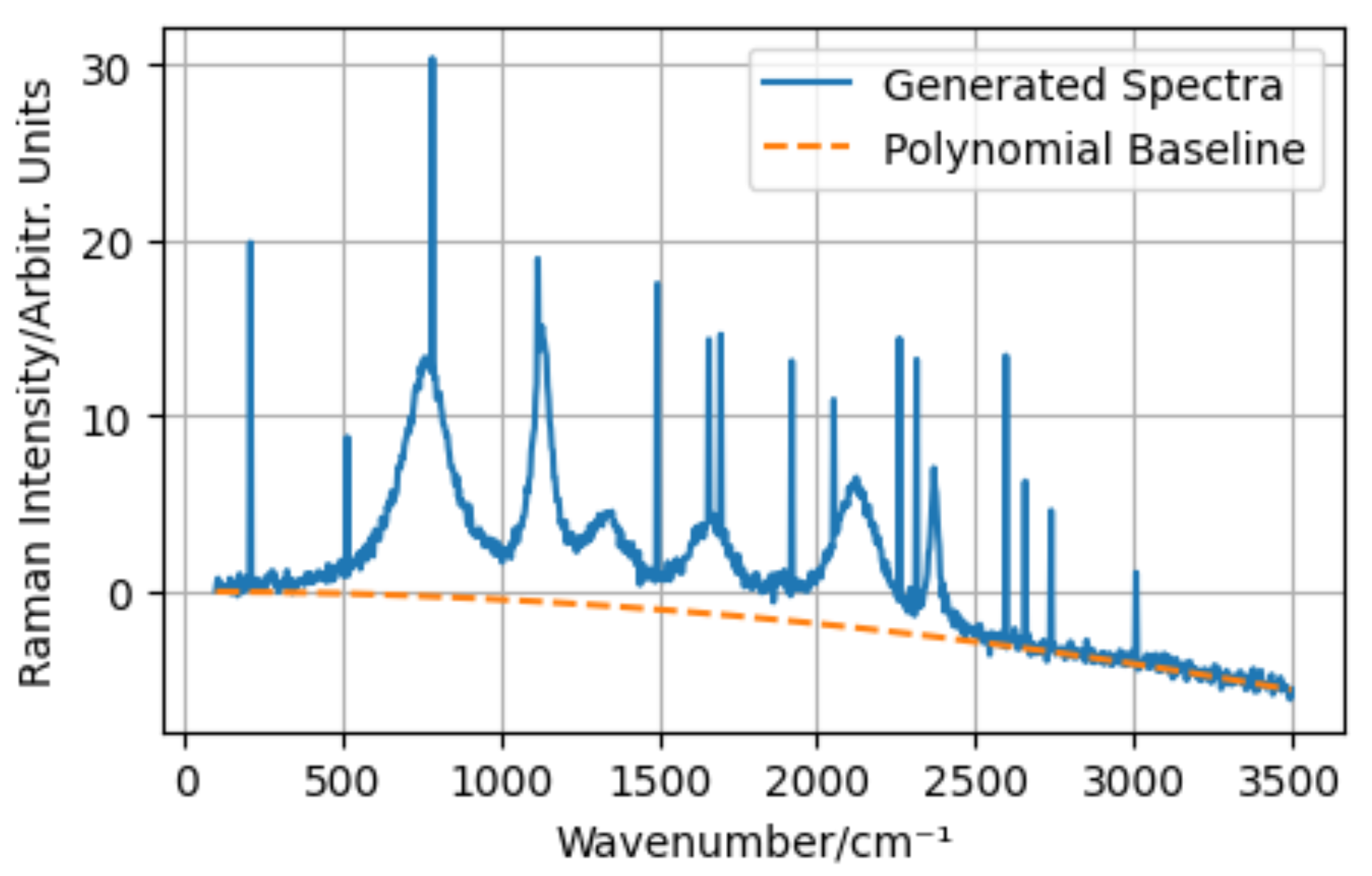}
\caption{
Example of a synthetically generated Raman spectrum composed of several components: simulated Voigt peaks (representing Raman signals), a polynomial fluorescence baseline, Gaussian-distributed noise, and high-amplitude spike artifacts mimicking cosmic rays. This dataset was used to validate the preprocessing and fitting procedures in a controlled setting.}
\label{fig:raman_generated}
\end{figure}

\section{Preprocessing}
The preprocessing of Raman spectra is a critical step that significantly influences the quality of the subsequent analysis \cite{guo2023key}. The primary objectives of preprocessing are to remove or reduce noise, correct baseline drift, and eliminate artifacts such as cosmic ray spikes. The steps involved in the preprocessing workflow are detailed below.
\subsection{Despiking
}

Cosmic ray suppression entails the removal of high-amplitude narrow spectral features that exceed the resolution limit of the spectrometer from the Raman spectrum. A highly effective method for this purpose is the modified Z-score algorithm. This algorithm quantifies the deviation of a value from the mean in terms of standard deviation. For a given Raman spectrum with mean $\mu$ and standard deviation $\sigma$, the Z-score for a given point $z(i)$ is calculated as follows:
\begin{equation}
\label{eq:zscore}
z(i) = \frac{x(i)-\mu}{\sigma}
\end{equation}

where $x(i)$ is the intensity value of the Raman spectrum in the $i$-th pixel (bin).
Z-scores can be substantially influenced by outliers. To improve resilience to such outliers, a modified Z-score method is used \cite{iglewicz1993volume}. The method substitutes the mean $\mu$ with the median $M$ and the standard deviation $\sigma$ with the median absolute deviation (MAD):

\begin{equation}
z(i) = 0.6745 \cdot \frac{x(i)-M}{MAD}
\end{equation}

\begin{figure}[h!]
  \centering
    \centering
    \includegraphics[width=\linewidth]{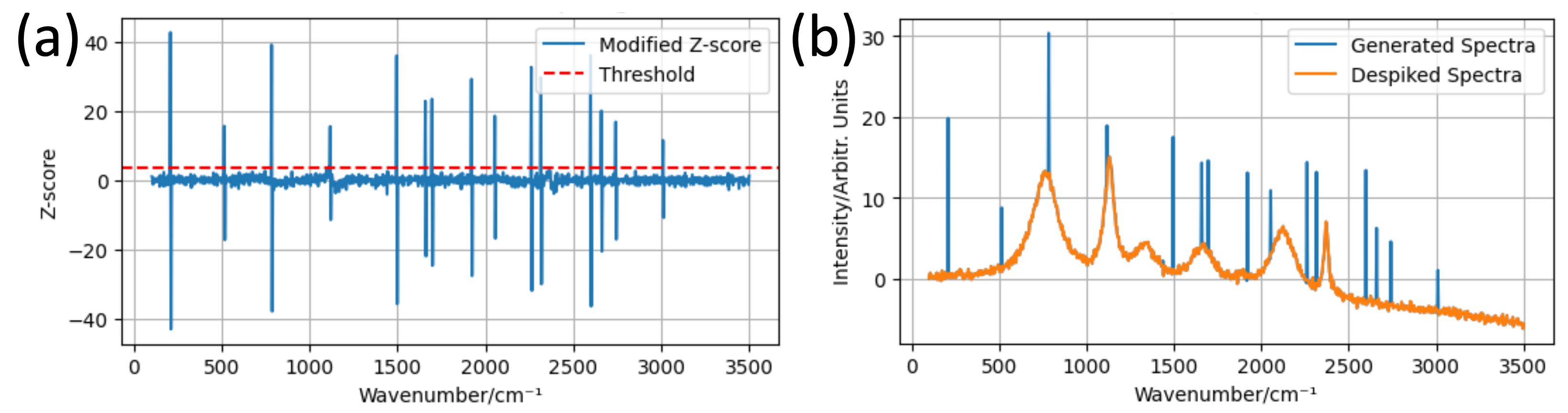}
  \caption{Cosmic ray spike suppression using the modified Z-score method. \textbf{(a)} Calculated modified Z-scores of first-order differences in the synthetic Raman spectrum. Points exceeding a threshold of 3.5 are identified as spikes. \textbf{(b)} Despiked Raman spectrum with spikes removed and missing values interpolated using neighboring points. This step improves peak fidelity and prepares the spectrum for further preprocessing.}
  \label{fig:combined_z}
\end{figure}

The constant 0.6745 adjusts the MAD to the standard deviation scale \cite{rodrigues2020tourist}.
To take advantage of the high intensity and small width of the spikes, the difference between consecutive spectrum points can be used to calculate the Z-score \cite{iglewicz1993volume}:

\begin{equation}
z(i) = 0.6745 \cdot \frac{\Delta x(i)-M}{MAD}
\end{equation}

where $\Delta x(i) = x(i) - x(i-1)$. The Z-score threshold is calibrated according to the dataset to attain the desired level of cosmic ray suppression. For the spectrum depicted in Fig. \ref{fig:raman_generated}, a threshold of 3.5 was selected, which is typically recommended by the American Society for Quality Control \cite{iglewicz1993volume}. Figure \ref{fig:combined_z} shows the calculated modified Z-scores and the resultant filtered spectrum, respectively.

The removal of cosmic ray signatures results in gaps within the spectrum. Interpolation methods are used to correct for these gaps by averaging neighboring values within a defined window. For the spectrum in Fig. \ref{fig:combined_z}, a window of $\pm8$ data points was used. The window was chosen in such a way that it could cover all gaps within the spectrum.

\subsection{Spectral Averaging}
To enhance the accuracy of Raman spectrum characterization, it is crucial to acquire multiple spectra of the sample at each sampled point whenever feasible. Averaging the recorded data is known to improve SNR proportionally to the square root of the number of replicates \cite{nakamura1989spectral}.

Before averaging, ensure that all Raman spectral datasets have consistent length, resolution, and wavenumber data points. In the case of discrepancies, interpolation methods such as piecewise linear interpolation \cite{blu2004linear} may be used to standardize the intensity values across the wavenumber data points. For given example we calculated the average of 10 generated Raman spectra with suppressed cosmic ray signatures, that showing consistent peaks and baselines but varying arbitrary noise.

When averaging multiple identical Raman spectra with Gaussian noise, the SNR will increase. This is because the signal, which is consistent across all spectra, will reinforce itself upon averaging, whereas the random noise, which varies, will partially cancel out.

After averaging \( n \) spectra, the averaged signal \( S_{\text{avg}} \) and the averaged noise \( N_{\text{avg}} \) can be expressed as:
\begin{equation}
S_{\text{avg}} = \frac{1}{n} \sum_{i=1}^{n} (S + N_i)
\end{equation}
Since \( S \) is constant across all spectra:

\begin{equation}
S_{\text{avg}} = S + \frac{1}{n} \sum_{i=1}^{n} N_i
\end{equation}

The noise in the averaged spectrum \( N_{\text{avg}} \) is:

\begin{equation}
N_{\text{avg}} = \frac{1}{n} \sum_{i=1}^{n} N_i
\end{equation}

The noise \( N_{\text{avg}} \) has a variance \( \sigma_{\text{avg}}^2 \) given by:

\begin{equation}
\sigma_{\text{avg}}^2 = \frac{\sigma^2}{n}
\end{equation}

Thus, the standard deviation of the noise in the averaged spectrum is:

\begin{equation}
\sigma_{\text{avg}} = \frac{\sigma}{\sqrt{n}}
\end{equation}

The SNR for a single spectrum is:

\begin{equation}
\text{SNR}_{\text{single}} = \frac{S}{\sigma}
\end{equation}

For the averaged spectrum, the SNR is:

\begin{equation}
\text{SNR}_{\text{avg}} = \frac{S_{\text{avg}}}{\sigma_{\text{avg}}} = \frac{S}{\sigma / \sqrt{n}} = \sqrt{n} \times \frac{S}{\sigma}
\end{equation}

The SNR of the averaged spectrum increases by a factor of \( \sqrt{n} \) compared to the SNR of a single spectrum. This means that by averaging 10 identical spectra with Gaussian noise, the SNR improves by a factor of:

\begin{equation}
\sqrt{10} \approx 3.16
\end{equation}

So, the SNR will \textbf{increase} by approximately 3.16 times when averaging 10 identical Raman spectra with Gaussian noise.

\subsection{Baseline Removal}
As previously noted, fluorescence in biological samples often overlaps with the Raman spectral region, complicating the differentiation between Raman features and the broad fluorescence baseline. To mitigate the impact of fluorescence on the spectrum, it is necessary to remove the baseline component associated with this fluorescence.

We employ an ALS fitting algorithm for this purpose. The parameters \( \lambda \) and \( p \) control the smoothness and asymmetry of the baseline, respectively.

In this context, \( p \) is the asymmetry parameter that determines the weighting between positive and negative residuals, with typical values ranging from 0.001 to 0.1. The parameter \( \lambda \) is the smoothing parameter, which controls the smoothness of the baseline with typical values ranging from \( 10^2 \) to \( 10^9 \).

The ALS algorithm iteratively fits the baseline by minimizing the following cost function:

\begin{equation}
\min \sum_{i} p \cdot \max(y_i - z_i, 0) + (1 - p) \cdot \max(z_i - y_i, 0) + \lambda \cdot \sum_{i} (\Delta^2 z_i)^2
\end{equation}

where \( y_i \) is the observed intensity, \( z_i \) is the estimated baseline, and \( \Delta^2 z_i \) represents the second-order difference of \( z_i \), ensuring smoothness.

For generated data, we used the following parameters: \( \lambda = 100,000,000 \) and \( p = 0.005 \), with 5 iterations. These values provided a good and smooth baseline.

To further enhance baseline correction, we evaluated an improved asymmetric reweighted penalized least squares (IARPLS) method \cite{ye2020baseline}, which builds on ALS by incorporating dynamic reweighting of residuals and optimized smoothing constraints. Unlike standard ALS, IARPLS adapts the asymmetry parameter \( p \) and smoothing parameter \( \lambda \) iteratively based on local spectral features, minimizing the influence of outliers and noise. This is achieved through the iterative formulation:

\[
\min \sum_{i} w_i \left( y_i - z_i \right)^2 + \lambda \cdot \sum_{i} (\Delta^2 z_i)^2
\]

where \( w_i \) are dynamic weights computed based on the residual \( r_i = y_i - z_i \), penalizing higher deviations to reduce the influence of noise.

IARPLS is more sensitive to the choice of its parameters compared to traditional ALS. Selecting optimal \( \lambda \) and \( p \) values is crucial, as improper tuning can lead to under- or over-fitting of the baseline, particularly in spectra with broad peaks or high background fluorescence.

Several automatic parameter tuning strategies can be applied to IARPLS. These include simple and iterative generate-evaluate approaches, high-level metaheuristics \cite{huang2020survey}, and cross-validation frameworks that maximize predictive correlation rather than minimize error \cite{holter2023r2}. In addition, the uniform design (UD) method has been proposed as a time-efficient alternative to traditional trial-and-error approaches \cite{wang2020uniform}. These methods enable more robust and data-adaptive parameter selection in IARPLS, enhancing reproducibility and generalization across datasets.

To further enhance baseline correction, we evaluated an improved asymmetric reweighted penalized least squares (IARPLS) method \cite{ye2020baseline}, which builds on ALS by incorporating dynamic reweighting of residuals and optimized smoothing constraints. Unlike standard ALS, IARPLS adapts the asymmetry parameter \( p \) and smoothing parameter \( \lambda \) iteratively based on local spectral features, minimizing the influence of outliers and noise. 

For our dataset, the IARPLS approach demonstrated superior performance in maintaining spectral features near the noise floor, particularly for samples with high fluorescence or overlapping peaks. Adaptive parameter tuning in IARPLS also required fewer iterations (typically 3) to converge to an optimal baseline.

\subsection{Smoothing}

In situations where replicates of the Raman spectrum at a single point on the sample cannot be obtained, low-pass filters may be utilized to enhance the SNR. However, this approach may lead to a loss of detailed information because of the transfer function's inability to provide a sharp cutoff at high spatial frequencies. Due to this limitation, it is essential to optimize the characteristics of the filter.

This technique can be applied in conjunction with spectral averaging to further reduce noise in Raman spectra. Various filtering methods can be employed to achieve this objective, including the Savitzky-Golay filter \cite{savitzky1964smoothing}, which is utilized in this study. The Savitzky-Golay filter is widely used for smoothing Raman spectra due to its ability to preserve the shape and position of spectral peaks while reducing signal noise. It operates by fitting adjacent data points in a sliding window with a polynomial using a linear least squares method.

Inappropriate selection of window width and polynomial order can result in data loss, often manifested as significant broadening of narrow lines and the merging of closely spaced spectral features.

For generated data, we used a window length of 11 and a polynomial order of 2.
The mathematical formula for the Savitzky-Golay filter is:

\begin{equation}
y'_i = \sum_{j=-k}^{k} c_j \cdot y_{i+j}
\end{equation}

where \( y'_i \) is the smoothed value, \( y_{i+j} \) are the original data points, \( c_j \) are the convolution coefficients, and \( k \) is half the window length.

The window size of the Savitzky-Golay filter determines the number of data points used in the local polynomial fitting process. It must be wide enough to capture the relevant trends in the signal without distorting the spectral features.

\begin{itemize}
    \item \textbf{Small window size:} Retains more of the sharp features of the spectrum (e.g., narrow Raman peaks) but may not remove noise effectively.
    \item \textbf{Large window size:} Smooths the noise more effectively but can cause loss of fine spectral details, leading to peak breadening or distortion.
\end{itemize}

They typical guidelines for selecting window size:
\begin{itemize}
    \item If the Raman spectrum contains narrow peaks, choose a smaller window size to avoid distorting them.
    \item If the spectrum has broad peaks or is noisy, a larger window size may be appropriate to smooth the data more.
    \item Typically, window sizes between 5 and 25 data points are commonly used in Raman spectroscopy. However, the choice should ultimately depend on the noise and resolution trade-offs in the data.
\end{itemize}

The polynomial order dictates how complex the fitted polynomial is for each window. The choice of polynomial order should be appropriate for the expected curvature of the spectral peaks.

\begin{itemize}
    \item \textbf{Low polynomial order:} Good for smooth signals or when the peaks are broad and don't require a lot of curvature fitting. Lower polynomial orders also reduce the risk of overfitting.
    \item \textbf{High polynomial order:} Can better model sharp, narrow peaks but risks fitting to noise and creating artificial features if overused.
\end{itemize}

For smooth spectra or when smoothing noise, it is better to use lower polynomial orders (e.g., 2 or 3). For spectra with sharp or complex peaks, higher polynomial orders (e.g., 4 or 5) may be more appropriate.

In most cases, visual inspection is sufficient to determine suitable filter parameters.

To determine the optimal parameters, an algorithm was implemented that systematically evaluates different combinations of window lengths and polynomial orders, calculating a score for each pair based on the balance between SNR and full width at half maximum (FWHM) of the spectral peaks. The score is defined as follows:

\begin{equation}
\text{Score} = \frac{\text{SNR}}{1 + \text{Relative FWHM Change}}
\end{equation}

where the relative FWHM change is calculated as the deviation of the FWHM from its baseline value before filtering. This approach ensures that the optimization process maximizes SNR while minimizing peak broadening, thereby preserving spectral features.

\begin{figure}[h!]
\centering
\includegraphics[width=0.6\linewidth]{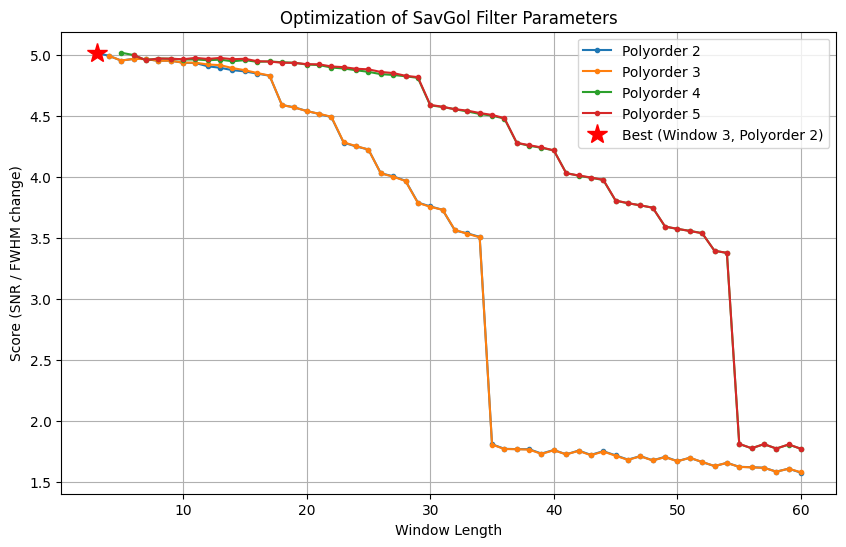}
\caption{Optimization of Savitzky-Golay filter parameters based on a custom score metric combining signal-to-noise ratio (SNR) and full width at half maximum (FWHM) preservation. The score is maximized when smoothing enhances SNR without significantly distorting peak shapes. The plot reveals that window lengths between 3–15 with polynomial order 2 achieve near-optimal balance, supporting robust spectral smoothing.}
\label{fig:savgol_optimization}
\end{figure}

Fig. \ref{fig:savgol_optimization} illustrates the relationship between window length and the computed score for different polynomial orders. The plot demonstrates the trade-off between smoothing and peak preservation for each combination of parameters. Based on the results, the optimal parameters for this dataset were determined to be a window length of 3 and a polynomial order of 2, as indicated by the highest score in the plot.

From this plot we can also see that the score does not change significantly in the window length range from 3 to 15, so any window length in this range is acceptable after visual examination of the plot after applying the Savitzky-Golay filter.

\section{Peak Analysis}

\subsection{Peak Detection}

Prior to fitting the spectrum, it is crucial to identify the peaks accurately, as correct peak identification enhances the precision and convergence of the fitting process. Numerous peak identification algorithms are available. The standard procedure for peak recognition typically involves the following steps:

\begin{itemize}
    \item \textbf{Detection of Local Maxima}: Peaks are identified by detecting local maxima in the Raman spectrum. A local maximum is defined as a point where the intensity exceeds that of its neighboring points.
    \begin{equation}
    I(x_i) > I(x_{i-1}) \quad \text{and} \quad I(x_i) > I(x_{i+1})
    \end{equation}
    where \( I(x) \) is the intensity at point \( x \).

    \item \textbf{Peak Height}: The height of each peak is determined by the intensity value at the local maximum, which provides an estimate of the peak amplitude \( A \).

    \item \textbf{Peak Width}: The width of each peak, often represented by the full width at half maximum (FWHM), is estimated by locating the points on either side of the peak where the intensity decreases to half of the peak height. For a peak centered at \( x_0 \) with height \( I(x_0) \), the FWHM is given by the distance between the points \( x_L \) and \( x_R \) where: 
    \begin{equation}
    I(x_L) = I(x_R) = \frac{I(x_0)}{2}
    \end{equation}

    \item \textbf{Peak Position}: The position of each peak is given by the wavenumber \( x \) at which the local maximum occurs. This wavenumber serves as the initial estimate for the peak center \( \mu \).
\end{itemize}

These initial estimates of peak parameters (position, height, and width) serve as the starting values for the Voigt fitting procedure. Utilizing these preliminary estimates facilitates a more efficient and rapid convergence of the fitting process.

\subsection{Voigt Peak Fitting}

The final step of the analysis involves fitting Voigt functions to the Raman spectrum. This fitting procedure is essential for accurately determining the positions, widths, and amplitudes of the peaks. Accuracy is particularly important in cases of overlapping peaks and for obtaining quantitative information about the analyzed sample.

\subsubsection{Voigt Function}
The Voigt function is defined as the convolution of a Lorentzian and a Gaussian function:

\begin{equation}
V(x, A, \mu, \sigma, \gamma) = A \cdot \int_{-\infty}^{\infty} \frac{\gamma}{\pi} \frac{1}{(x' - \mu)^2 + \gamma^2} \cdot \frac{1}{\sigma \sqrt{2\pi}} e^{-\frac{(x - x')^2}{2\sigma^2}} dx'
\end{equation}

where \( A \) is the amplitude, \( \mu \) is the center (or position) of the peak, \( \sigma \) is the standard deviation of the Gaussian component, and \( \gamma \) is the Lorentzian (natural) half-width at half-maximum (HWHM).

Alternatively, the Voigt function can be expressed using the real part of the Faddeeva function \( w(z) \):

\begin{equation}
V(x, A, \mu, \sigma, \gamma) = A \cdot \frac{\operatorname{Re}\left[ w\left( \frac{x - \mu + i \gamma}{\sigma \sqrt{2}} \right) \right]}{\sigma \sqrt{2\pi}}
\end{equation}

\subsubsection{Composite Function}
To fit multiple Voigt peaks, we define a composite function as the sum of individual Voigt functions:

\begin{equation}
f(x, \{A_i, \mu_i, \sigma_i, \gamma_i\}) = \sum_{i} V(x, A_i, \mu_i, \sigma_i, \gamma_i)
\end{equation}

where \( \{A_i, \mu_i, \sigma_i, \gamma_i\} \) are the parameters for each Voigt peak.

\subsubsection{Fitting Procedure}
The fitting process aims to minimize the discrepancy between the observed spectrum and the fitted curve by adjusting the parameters of the Voigt functions. A common approach is to use local least squares minimization to refine the fit. This method fine-tunes the peak parameters to minimize the sum of the squares of the residuals between the observed data and the model.

In cases where the number of parameters is large or the initial estimates are not close to the optimal solution, a global optimization method, such as differential evolution \cite{storn1997differential}, can be employed to obtain robust initial estimates of the peak parameters. However, for most practical purposes, local optimization methods are sufficient and more computationally efficient.

Additionally, the fitting can be performed iteratively by analyzing the residuals from the previous fit to identify any peaks that may have been missed initially. After the initial fitting, the residuals (the difference between the observed spectrum and the fitted model) can be examined for significant features that indicate the presence of additional peaks. These residual peaks can then be included in a subsequent fitting iteration. By iteratively adding peaks based on the residuals, the fitting process can progressively improve the accuracy of the model.

An example of this iterative fitting approach is implemented in our code, where after each fitting iteration, the residuals are analyzed using peak detection algorithms to find any remaining peaks. These peaks are then added to the model, and the fitting is repeated until no significant peaks remain in the residuals or a maximum number of iterations is reached. We employ an iterative fitting approach that involves analyzing the residuals after each fitting iteration to identify any peaks that may have been missed initially.

The fitting procedure consists of the following steps:

\begin{enumerate}
    \item \textbf{Initial Peak Detection}: Detect peaks in the pre-processed Raman spectrum using a peak detection algorithm. This provides initial estimates for the peak positions, heights, and widths.
    \item \textbf{Initial Fitting}: Fit the Raman spectrum using a composite model comprising Voigt functions for each detected peak. The fitting is performed using local least squares minimization to optimize the parameters of the Voigt functions.
    \item \textbf{Residual Analysis}: Calculate the residuals by subtracting the fitted model from the observed spectrum. Analyze the residuals using a peak detection algorithm to identify any additional peaks that were not accounted for in the initial fitting.
    \item \textbf{Iterative Refinement}: If significant peaks are found in the residuals, add Voigt functions corresponding to these peaks to the composite model. Repeat the fitting process using the updated model. This iterative process continues until no significant peaks remain in the residuals or a predefined maximum number of iterations is reached.
    \item \textbf{Model Simplification}: After each fitting iteration, evaluate the amplitudes of the fitted peaks. Peaks with negligible amplitudes (below a certain threshold) are considered insignificant and are removed from the model to prevent overfitting.
    \item \textbf{Final Model}: The final fitted model consists of the sum of Voigt functions that best represent the spectral features of the Raman spectrum. The parameters of the Voigt functions provide accurate estimates of the peak positions, widths, and amplitudes.
\end{enumerate}

The overall objective function to be minimized is:

\begin{equation}
\min_{\{A_i, \mu_i, \sigma_i, \gamma_i\}} \sum_{j} \left( y_j - f(x_j, \{A_i, \mu_i, \sigma_i, \gamma_i\}) \right)^2
\end{equation}

where \( y_j \) is the observed intensity at wavenumber \( x_j \), and \( f(x_j, \{A_i, \mu_i, \sigma_i, \gamma_i\}) \) is the composite Voigt model.

This iterative fitting approach allows for the progressive refinement of the model by accounting for peaks that may not be apparent in the initial spectrum due to overlapping features or low SNR. By continuously updating the model based on the residuals, we enhance the accuracy of peak characterization and ensure that all significant spectral features are captured.

An example of this iterative fitting process is implemented in our code, where after each fitting iteration, the residuals are analyzed using peak detection algorithms to find any remaining peaks. These peaks are then added to the model, and the fitting is repeated until convergence is achieved.

\begin{figure}[h!]
\centering
\includegraphics[width=1\linewidth]{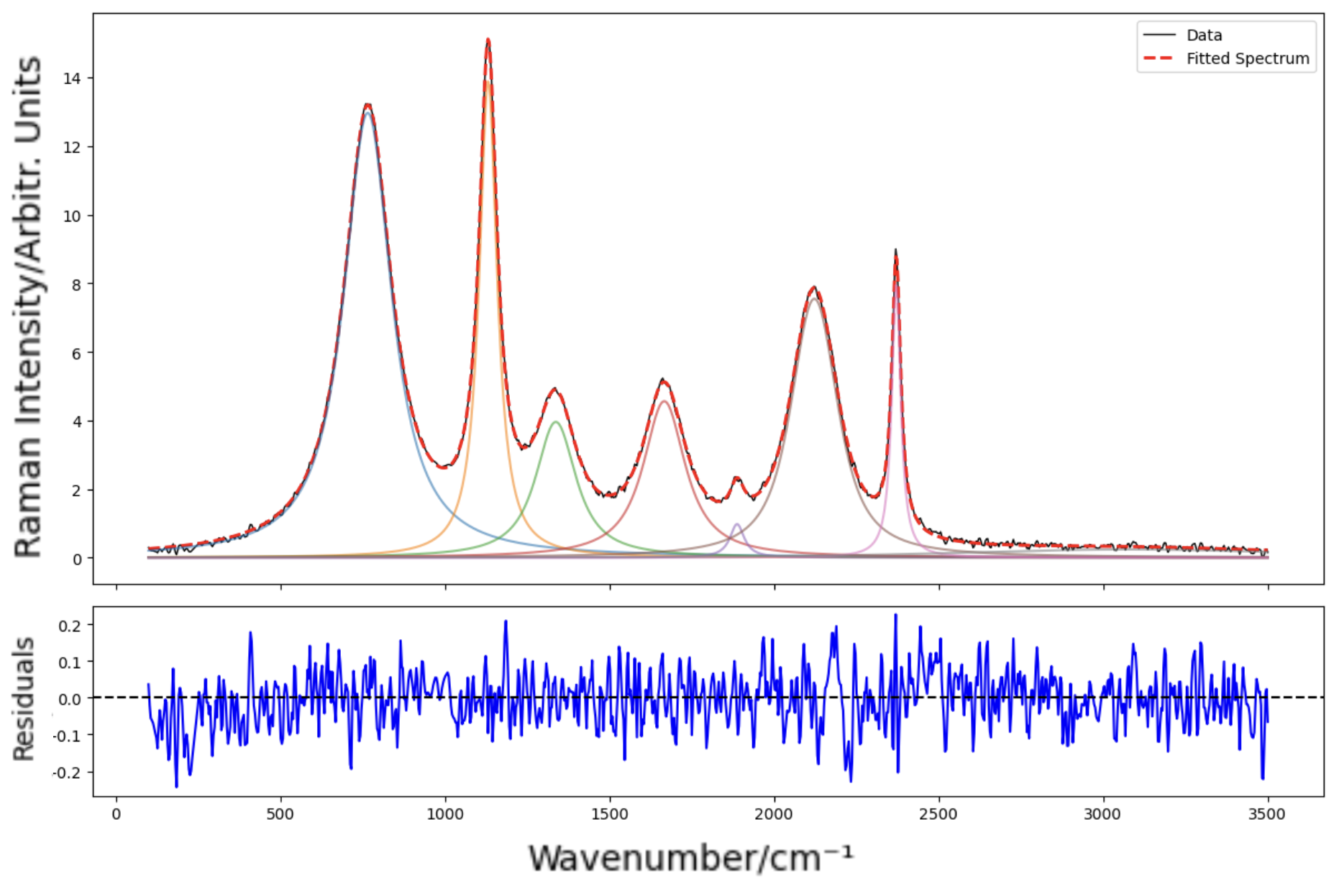}
\caption{Final Voigt peak fitting results for the synthetic Raman spectrum. The fitted model accurately captures all simulated peaks, including overlapping ones, using iterative residual analysis and model refinement. Residuals between the data and the fitted model (shown below) confirm excellent agreement and support the utility of Voigt-based modeling in complex spectral scenarios.}
\label{fig:raman_generated_fit}
\end{figure}

The final fitted curve, its residuals, and the individual Voigt peaks are illustrated in Fig. \ref{fig:raman_generated_fit}. This fitting method ensures precise peak identification, which is crucial for the accurate analysis and interpretation of Raman spectra.

\section{Real-World Validation}
\label{real_data}

In this section, we demonstrate the application of the Raman spectrum data preprocessing and analysis pipeline on real-world data acquired using custom-built spectrometers. The goal of these analyses is not only to extract meaningful spectral information but also to evaluate the robustness and versatility of the proposed pipeline across different experimental setups and sample types. By applying the algorithm to distinct datasets, we aim to validate its reliability, reproducibility, and adaptability in various real-world conditions. This multi-system validation is essential for confirming the generalizability of our workflow beyond synthetic or controlled datasets.

Each experimental setup introduces unique challenges, including variations in instrumentation, excitation wavelengths, SNR, and sample characteristics. For example, Raman spectra collected using deep-ultraviolet excitation may exhibit distinct fluorescence backgrounds or noise artifacts compared to spectra collected using visible light excitation. Similarly, different detectors and spectrometers present varying levels of sensitivity, resolution, and noise patterns. Demonstrating the effectiveness of our preprocessing and analysis techniques across these varied conditions highlights the robustness and scalability of the workflow.

\subsection{Custom-Built Microspectrometers}
\begin{figure}
    \centering
    \includegraphics[width=0.9\linewidth]{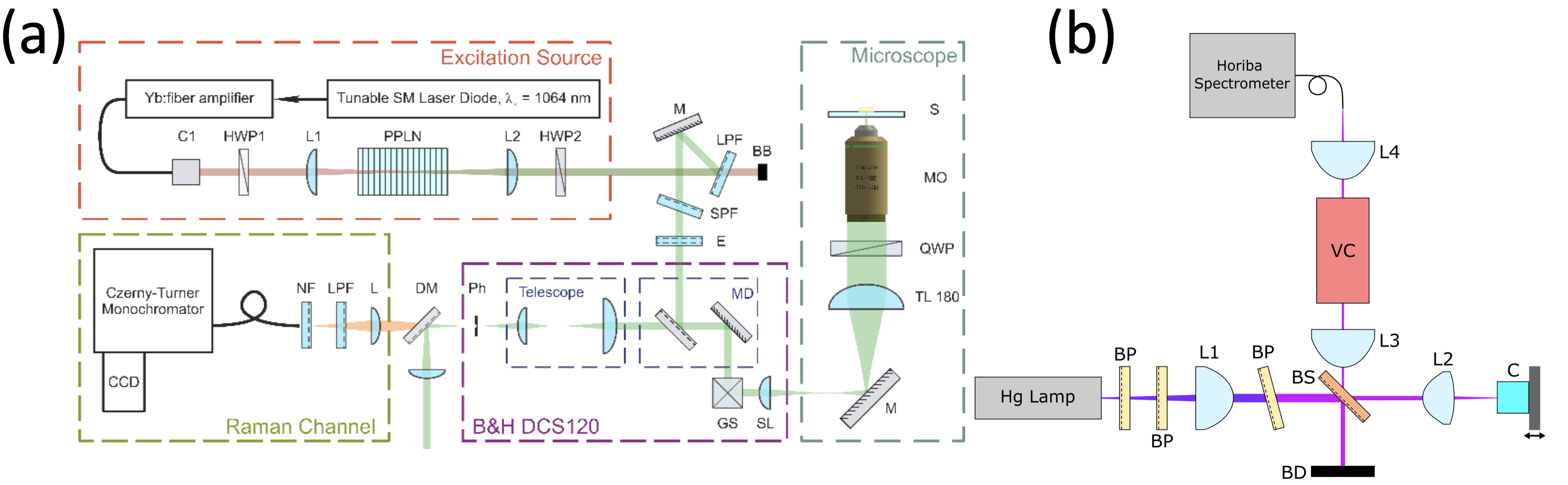}
    \caption{Schematics of two custom-built Raman systems used in this study. \textbf{(a)} Confocal Raman microspectrometer with 532\,nm excitation optimized for high-resolution point measurements. \textbf{(b)} Deep-UV resonant Raman (DUVRR) system using \SI{254}{\nano\meter} excitation and a mercury vapor cell for Rayleigh suppression. Both systems were used to validate the workflow on real-world Raman spectra from diverse sample types. Confocal Raman microspectrometer components: \textbf{C} – fiber collimator, \textbf{HWP} – half-wave plate, \textbf{L} – plano-convex lens, \textbf{M} – mirror, \textbf{LPF} – long-pass hard-coated interference filter, \textbf{SPF} – short-pass hard-coated interference filter, \textbf{BB} – beam block, \textbf{E} – fused silica Fabry-Pérot etalon, \textbf{MD} – main dichroic assembly, \textbf{GS} – galvo-galvo scanner, \textbf{SL} – scan lens, \textbf{TL} – tube lens, \textbf{QWP} – quarter-wave plate, \textbf{MO} – microscope objective lens, \textbf{S} – sample, \textbf{Ph} – precision pinhole wheel, \textbf{DM} – dichroic mirror, \textbf{NF} – notch filter, \textbf{CCD} – camera with a charge-coupled device sensor. DUVRR components: \textbf{BP} – Bandpass filter, \textbf{L} – aspheric lens, \textbf{BD} – beam dump, \textbf{BD} – 50/50 beamsplitter, \textbf{C} – cuvette, \textbf{VC} – Hg atomic vapor cell.}
    \label{fig:microspectrometers}
\end{figure}

In this section, we describe two custom-built Raman systems used to obtain real-world data for evaluating our preprocessing and analysis workflow. The first setup is a confocal Raman microspectrometer operating in the visible range (532\,nm excitation), and the second is a deep-UV resonant Raman (DUVRR) system operating at approximately 254\,nm excitation. Both systems are designed to measure Raman spectra with minimal fluorescence and to acquire sufficient signal with low laser power. A schematic of the systems used to acquire data is shown in Fig. \ref{fig:microspectrometers}.

\subsubsection{Custom-Built Confocal Raman Microspectrometer}

The microspectrometer was built around a tunable single longitudinal mode laser diode (Koheras Adjustik Y10, NKT Photonics) with a spectral linewidth of less than 1 MHz. This laser was coupled to an ytterbium-doped fiber amplifier (Koheras Boostik HPA, NKT Photonics), which provided an output of 1.8 W at 1064 nm.

A second harmonic at 532 nm was generated using a periodically poled lithium niobate (PPLN) crystal (Covesion Ltd.), as indicated in Fig. \ref{fig:microspectrometers}. Residual 1064 nm radiation was separated from the 532 nm output using a combination of two interference filters, which together achieved a total optical density greater than 6.0 (Semrock). 

The excitation beam was directed into the input port of a DCS120 confocal scan head (Becker \& Hickl), which was mounted on a custom-built inverted microscope. Infinite conjugate microscope objectives were utilized for both delivering the excitation light and collecting the scattered photons. Infinite conjugate optics were employed to facilitate the installation of beamsplitters and filters behind the microscope objective without requiring refocusing. Consequently, a 180 mm focal length tube lens was incorporated to couple the light to and from the scan head. The typical power delivered to the samples did not exceed \SI{5}{\milli\watt}.

The collected signal was routed to a Czerny-Turner spectrometer (Shamrock 303, Oxford Instruments) through a circle-to-line fiber bundle with a numerical aperture (NA) of 0.12 and a core diameter of 160 $\mu$m. Elastically scattered photons were isolated using a combination of a notch and long-pass filter. The total optical density of the filter assembly exceeded 60 dB.

The CCD camera in the spectrometer assembly was operated in full vertical binning (FVB) mode. To minimize thermal noise, the camera was cooled to -70 $^{\circ}$C and maintained at this temperature. Acquisition times for each spectrum ranged from 5 to 8 minutes, depending on the visible SNR and the perceived efficiency of Raman scattering in the samples. The mechanical slit was kept open, with the fiber core serving as the entrance slit for the spectrometer. Each sample was interrogated a minimum of three times to acquire multiple Raman spectra, thereby enhancing the SNR and improving the robustness of the analysis.
\subsubsection{Custom-Built Deep-UV Resonant Raman}

The Raman spectra were collected using a custom-built deep-UV resonant Raman system designed for efficient signal enhancement and low-wavenumber exploration. The excitation source was a low-pressure mercury lamp (Jelight Company, Inc.), which passed through a series of \qty{254}{\nano\meter} MaxLamp\textsuperscript{\texttrademark} mercury line filters (Semrock, Inc.) and Schott UG5 glass to isolate the \qty{253.65}{\nano\meter} excitation line. The UV light is then collimated with a DUV fused silica aspheric lens (Thorlabs, Inc.), after which a rotated, blue-shifted \qty{266}{\nano\meter} MaxLine\textsuperscript{\texttrademark} laser clean-up filter (Semrock, Inc.) is used to remove an inherent ``shoulder'' on the Stokes side of the \qty{253.65}{\nano\meter} excitation line.

The filtered light then transmits through a 50/50 beamsplitter (Thorlabs, Inc.) to a sample arm. A UV AR-coated aspheric lens (Edmund Optics, Inc.) was used for both excitation and Raman signal collection in a backscattered geometry. The collected signal was then reflected off of the 50/50 beamsplitter and through an aspheric lens (Thorlabs, Inc.) and into a fused silica mercury vapor cell (Opthos Instrument Company, LLC.) heated to an upper limit of \qty{200}{\celsius}. The vapor cell suppresses the Rayleigh scattering of the sample by more than two orders of magnitude, allowing for potential exploration of the low-wavenumber of Raman shifts. 

After elastic scattering suppression, an aspheric lens (Thorlabs, Inc.) coupled the Raman signal into a round-to-linear fiber bundle (FiberTech Optica, Inc.), which was connected to a Horiba iHR 320 spectrometer equipped with a liquid nitrogen-cooled CCD detector. The spectrometer utilized a 2400 grooves/mm grating with a \qty{330}{\nano\meter} blaze wavelength, optimized for high efficiency in the UV spectral regime. 

The spectrometer settings, including slit width, acquisition time, and the number of spectral runs, were adjusted depending on the specific sample and experimental requirements. These parameters will be detailed in the relevant sections discussing individual sample analyses. The incoherent nature of the mercury lamp resulted in a relatively loose focus at the sample, which is beneficial for future wide-field, speckle-free imaging experiments. To maximize photon collection efficiency, a high numerical aperture (NA = 0.65) lens was employed. Due to the resonant enhancement of the system, samples were never exposed to greater than \qty{10}{\micro\watt} of power \cite{jhraman, 2025HarringtonDUVSPIE}.

\subsection{Raman Analysis of Real Data}

To demonstrate the robustness of our workflow on experimentally acquired datasets, we analyzed three different samples: Teflon (PTFE), a medical ointment, and polymeric orthodontics aligners (PET-G). The data were collected from the two custom-built systems detailed earlier: (1) a confocal Raman microspectrometer operating at 532\,nm for the ointment sample, and (2) a deep-UV resonant Raman (DUVRR) system at approximately 254\,nm for Teflon and aligners. In all cases, we applied the same preprocessing pipeline that shown on Fig. \ref{fig:raman_workflow}.

\subsubsection{Medical Ointment}

\begin{figure}[h!]
    \centering
    \includegraphics[width=0.8\linewidth]{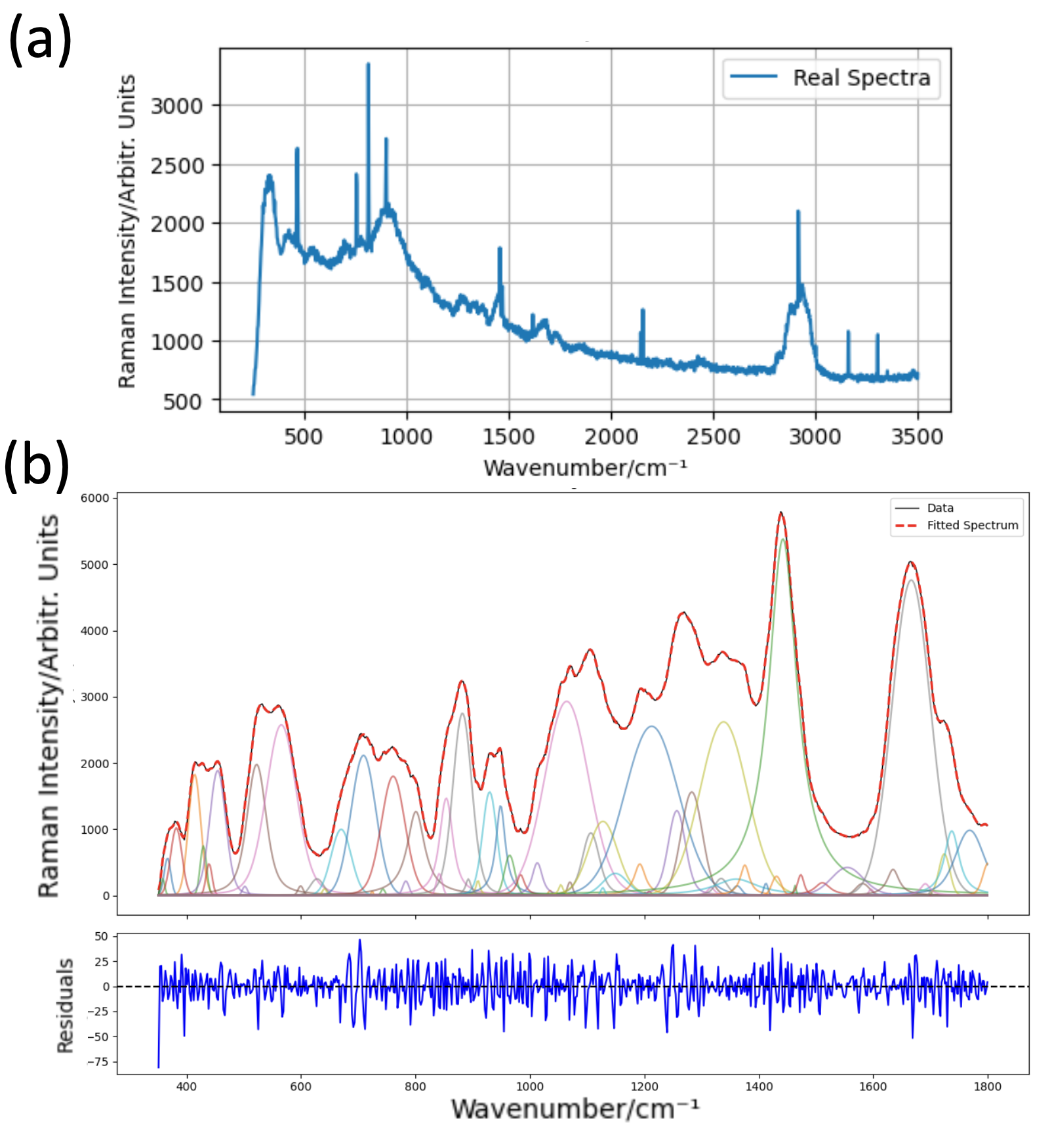}
    \caption{
    \textbf{(a)} Raw Raman spectrum of a medical ointment acquired using 532\,nm excitation. Fluorescence background and spike artifacts are visible. 
    \textbf{(b)} Processed spectrum after cosmic ray suppression, baseline correction, and Voigt peak fitting. This example illustrates the full preprocessing and fitting workflow applied to real-world data.
    }
    \label{fig:raman_med}
\end{figure}

We investigated a medical ointment sample on the confocal microspectrometer at \SI{532}{\nano\meter} excitation. As shown in Fig.~\ref{fig:raman_med}, the raw spectrum contained noticeable fluorescence background and random spikes. Our standard pipeline effectively mitigated these artifacts, yielding a clean, baseline-corrected spectrum amenable to Voigt-peak fitting. Table~\ref{table:table} lists the primary peaks identified in the spectrum, together with their functional groups and usage. Notably, characteristic bands from aromatic compounds, alkanes, sulfur-containing drugs, and lipid structures were readily discernible, reflecting the complex formulation typical of medical ointments.

As shown, the residual closely overlaps with the original spectrum, indicating a good fit; however, many fitted peaks in this region are not scientifically meaningful \cite{bradley2007curve}.

Note on the broad peak near 900 cm$^{-1}$: In Fig.~\ref{fig:raman_med}(a), a prominent broad feature appears near 900~cm$^{-1}$ in the raw spectrum. This feature is absent in the corrected spectrum shown in Fig.~\ref{fig:raman_med}(b). It is important to emphasize that this feature is primarily due to strong background fluorescence rather than Raman scattering. To eliminate this effect, we recorded a separate background spectrum (with no sample present) under identical acquisition conditions and subtracted it directly from the raw sample spectrum. This physically motivated subtraction step effectively removed the broad fluorescent background, including the 900~cm$^{-1}$ feature. Since this is not a Raman peak, its absence in the corrected spectrum confirms the success of the experimental background removal rather than indicating an artifact introduced by the workflow.

After fitting, we can obtain the peak wavenumber and intensity values, which allow us to analyze the structure of the imaged ointment \cite{lambert1987introduction}.
The peaks identified in the Raman spectrum (see Table \ref{table:table}) align well with what is expected from a typical medical ointment. Below is a brief discussion of the relevance of several most important detected peaks:

\begin{itemize}
    \item \textbf{Aromatic Compounds (C-H bending and out-of-plane bending)}: 
    Aromatic rings are common in many pharmaceutical agents or preservatives used in ointments, contributing to peaks around 394 cm$^{-1}$, 700 cm$^{-1}$, and 774 cm$^{-1}$.

    \item \textbf{Alkanes, Alcohols (C-C stretching, C-H wagging)}: 
    Alkanes and alcohol-based components are often present in ointments as oils, waxes, and solvents. The peaks around 400 cm$^{-1}$, 453 cm$^{-1}$, and 1270 cm$^{-1}$ correspond to these structures.

    \item \textbf{Sulfur-containing Drugs (C-S stretching)}: 
    The peaks around 521 cm$^{-1}$ and 564 cm$^{-1}$ suggest the presence of sulfur-based compounds, which are common in antibacterial and antifungal medications.

    \item \textbf{Water and Alcohols (O-H bending)}: 
    The peak at 875 cm$^{-1}$ is likely due to O-H bending, indicating the presence of alcohols or water, which are common excipients in moisturizing or soothing ointments.

    \item \textbf{Esters, Ethers, Lipids, and Fats (C-O stretching, CH$_2$ twisting)}: 
    Peaks around 1056 cm$^{-1}$ and 1351 cm$^{-1}$ correspond to esters, ethers, and lipid structures, which are often used as emollients to soften and protect the skin.

    \item \textbf{Ketones, Esters, and Fatty Acids (C=O stretching)}: 
    The peak at 1659 cm$^{-1}$ indicates the presence of carbonyl groups, which are typical in ketones, esters, and fatty acids—ingredients commonly used as carriers or active components in topical medications.
\end{itemize}

The identified peaks correspond well to a complex mixture containing active ingredients, solvents, moisturizers, and stabilizers—common components in medical ointments. This suggests that the Raman spectrum obtained provides a reliable insight into the molecular composition of the ointment and aligns with its expected chemical makeup.

\begin{table}[h!]
\centering
\caption{Matched Raman Peaks with Functional Groups}
\begin{tabular}{ccc}
\toprule
\textbf{Peak Center (cm$^{-1}$)} & \textbf{Functional Group / Molecule} & \textbf{Common Usage} \\ 
\midrule
394.62 & C-H bending (aromatic) & Aromatic compounds \\ 
400.09 & C-C stretching (aliphatic) & Alkanes, alcohols \\ 
419.95 & C-C stretching (aliphatic) & Alkanes, alcohols \\ 
453.49 & C-C stretching (aliphatic) & Alkanes, alcohols \\ 
521.54 & C-S stretching & Sulfur-containing drugs \\ 
564.86 & C-S stretching & Sulfur-containing drugs \\ 
700.45 & C-H out-of-plane bending & Aromatic compounds, resins \\ 
731.05 & C-H out-of-plane bending & Aromatic compounds, resins \\ 
774.66 & C-H out-of-plane bending & Aromatic compounds, resins \\ 
875.78 & O-H bending & Alcohols, water in ointments \\ 
1056.07 & C-O stretching & Esters, ethers, alcohols \\ 
1270.14 & C-H wagging & Alkanes \\ 
1351.98 & CH$_2$ twisting & Lipids, fats \\ 
1441.28 & C-H bending & Alkyl groups, proteins \\ 
1659.19 & C=O stretching & Ketones, esters, fatty acids \\ 
\bottomrule
\label{table:table}
\end{tabular}
\end{table}

\subsubsection{PTFE}

We analyze the Raman peaks obtained from the Teflon sample using UV Raman spectroscopy and compare them with the well-established Raman peaks characteristic of PTFE, commonly known as Teflon. For Teflon Raman peaks are well-documented in the literature, corresponding to its characteristic molecular vibrations.

\begin{figure}[h!]
        \centering
        \includegraphics[width=0.8\linewidth]{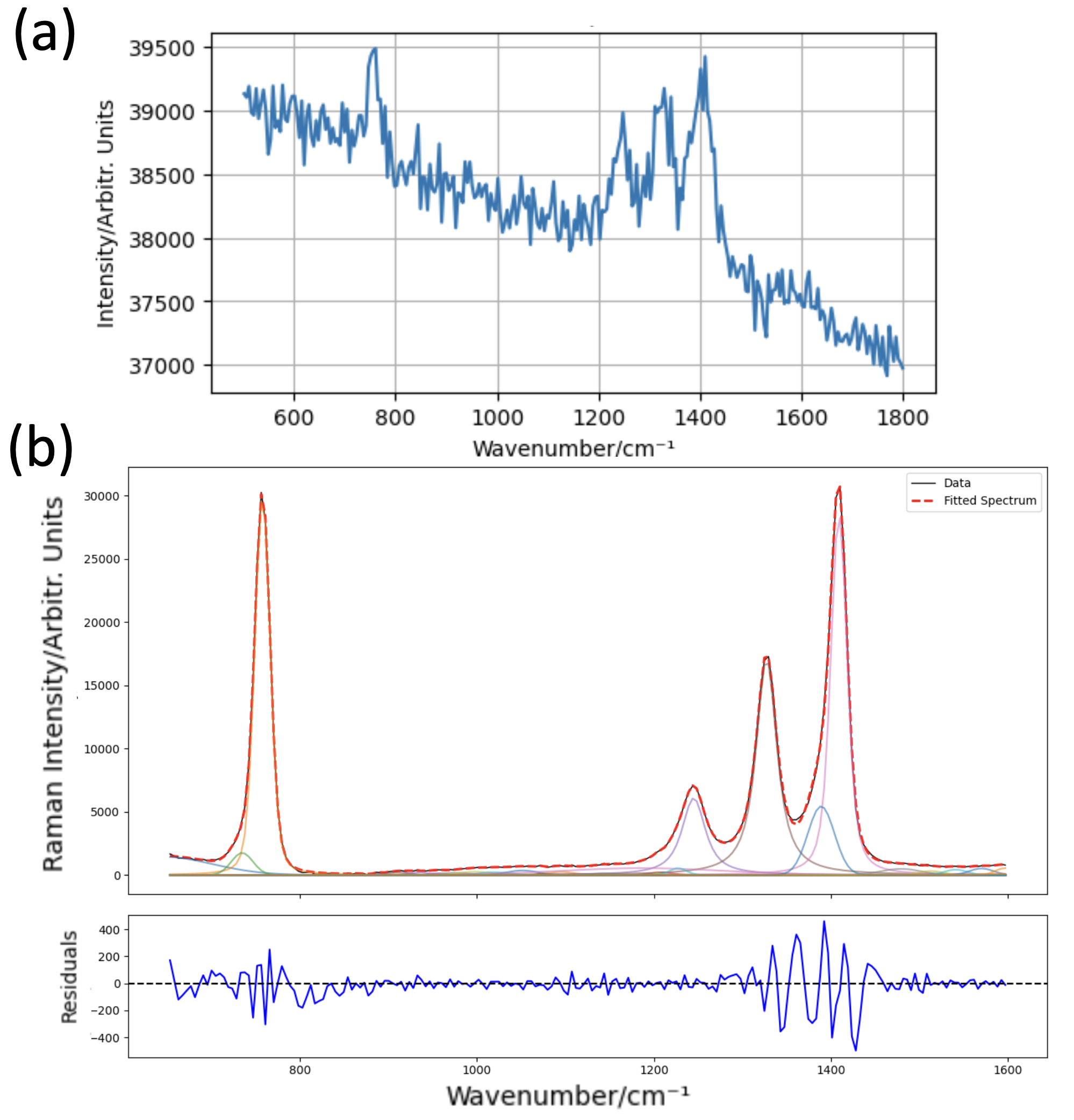}
        \caption{Raman spectra of Teflon (PTFE) acquired using the DUVRR system. \textbf{(a)} Raw spectrum showing significant noise and weak signal. \textbf{(b)} Cleaned and fitted spectrum following despiking, baseline correction, and smoothing. Characteristic PTFE peaks are resolved and modeled with Voigt profiles. This analysis validates the framework on low-SNR industrial polymer samples.}
        \label{fig:raman_teflon}
\end{figure}

Raman spectra of Teflon were initially very noisy and thus difficult to interpret, as seen in Fig.~\ref{fig:raman_teflon}. But after preprocessing we managed to obtain peak information.

For this sample, Raman spectra were acquired using a slit width of \qty{100}{\micro\meter} and an acquisition time of \qty{120}{\second} per run. To enhance the signal-to-noise ratio, four consecutive measurements were collected and averaged. A high NA (0.65) lens was used to maximize the collection of scattered photons, and the excitation power at the sample was kept around \qty{2}{\micro\watt}.

The Raman peaks for Teflon are typically observed at the following wavenumbers: 731 cm$^{-1}$ (C-C stretching), 1210 cm$^{-1}$ (CF$_2$ symmetric stretching), 1295 cm$^{-1}$ (CF$_2$ asymmetric stretching), and 1380 cm$^{-1}$ (C-F bending) \cite{koenig1969raman}. These peaks serve as spectral markers for the presence of Teflon, allowing for precise identification in Raman spectra.

In the current experiment, after preprocessing and fitting Raman spectra (see Fig. \ref{fig:raman_teflon}) the following Raman peaks were measured from the sample (see Table \ref{tab:summary_teflon_peaks}). The observed peaks show good agreement with the expected Teflon peaks, confirming the presence of Teflon in the analyzed sample. Notably, peaks near 740 cm$^{-1}$ and 1384 cm$^{-1}$ align closely with the expected positions for C-C and C-F vibrations, respectively.

Although several Raman peaks in the PTFE spectrum were systematically shifted by 10–20 $cm^{-1}$ compared to literature values, this can be attributed to UV excitation effects, instrumental calibration differences, and Voigt-based deconvolution. The band at 1385 $cm^{-1}$ corresponds to the asymmetric C–F bending mode, consistent with literature reports near 1379 $cm^{-1}$, albeit shifted slightly due to the above factors

\begin{table}[h!]
\centering
\caption{Summary of Measured Peaks and Comparison with Known Teflon Peaks.}
\begin{tabular}{ccc}
\toprule
\textbf{Known Teflon Peak (cm$^{-1}$)}  & \textbf{Amplitude} \\
\midrule
731   & 307817 \\
1210 & 355852 \\
1295 & 741806 \\
1380 & 1059809 \\
\bottomrule
\end{tabular}
\label{tab:summary_teflon_peaks}
\end{table}

The excellent match between the measured and expected Raman peaks supports the conclusion that the analyzed material is consistent with Teflon. Minor deviations in peak positions (e.g., 740.99 cm$^{-1}$ instead of 731 cm$^{-1}$) could result from experimental factors such as temperature fluctuations, sample preparation methods, or instrumental resolution. Additionally, the observed peak widths (FWHM) reflect the sample's heterogeneity and measurement conditions.

The Raman peaks observed at 1243 cm$^{-1}$ and 1384 cm$^{-1}$ correspond to the characteristic CF$_2$ symmetric stretching and C-F bending vibrations, further validating the sample's identity as Teflon. The alignment of these peaks with known Teflon spectra demonstrates the reliability of the experimental setup and the robustness of Raman spectroscopy for material identification.

\subsubsection{Polymeric Aligners
}
\begin{figure}[h!]
    \centering
    \includegraphics[width=0.8\linewidth]{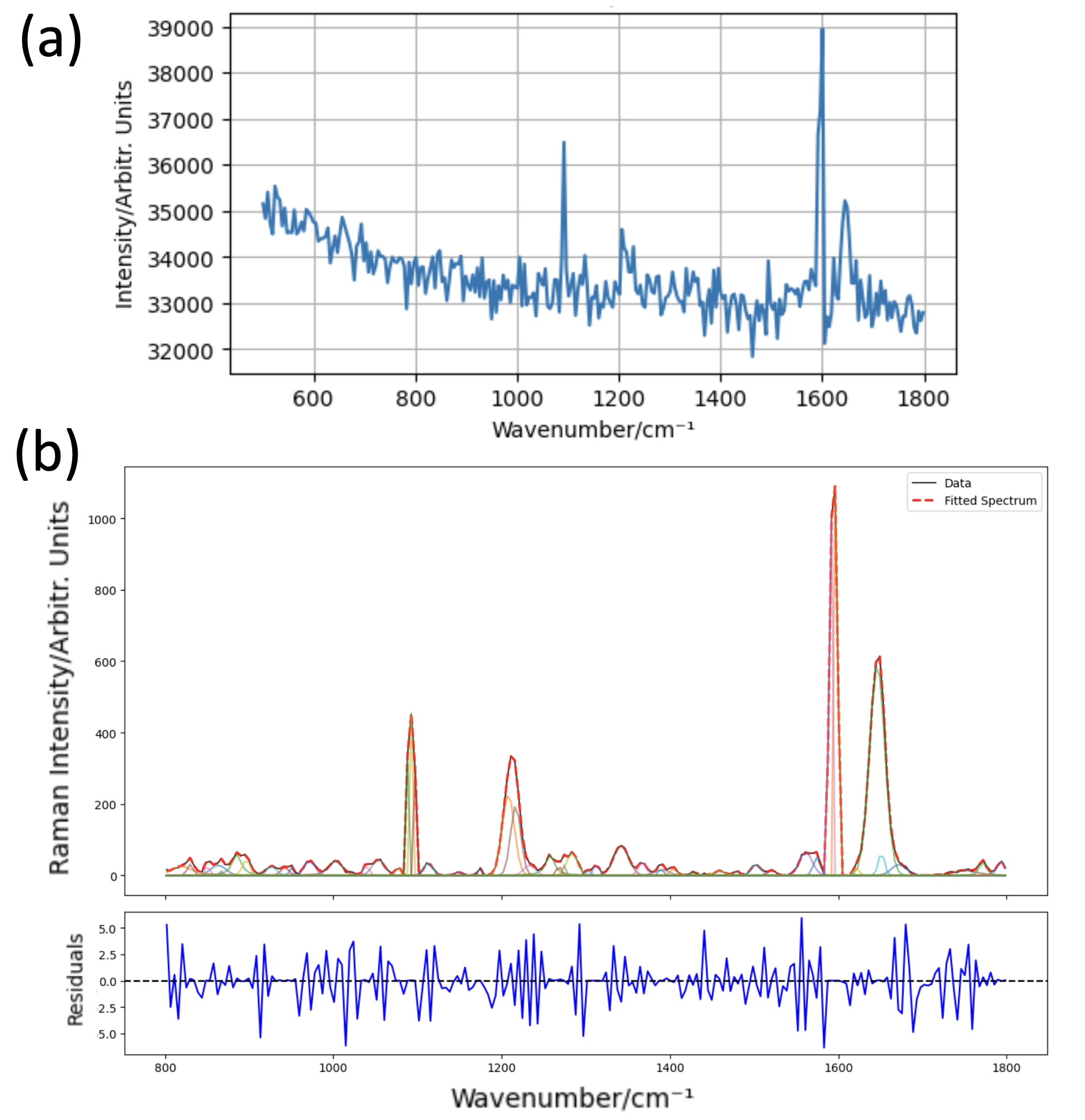}
    \caption{Raman spectra of polymeric orthodontic aligners (PET-G). \textbf{(a)} Raw Raman signal containing strong noise and minor baseline drift. \textbf{(b)} Spectrum after full preprocessing and Voigt peak fitting. Several known PET-G vibrational modes (e.g., C=O and C-H stretching bands) were identified, confirming the robustness of the workflow for complex polymers under low-power UV excitation.}
    \label{fig:raman_invisalign}
\end{figure}

The initial Raman data for orthodontics aligners was similarly noisy and unsuitable for direct interpretation, as seen in Fig.~\ref{fig:raman_invisalign}. However, through our iterative preprocessing strategy, we successfully suppressed noise and extracted meaningful peak information.

Data was collected using a \qty{100}{\micro\meter} slit width and an acquisition time of \qty{120}{second} per run. Each spectrum was averaged over four consecutive acquisitions to improve the signal-to-noise ratio. Additionally, a high numerical aperture (NA = 0.65) lens was utilized to maximize photon collection, while the excitation power remained below \qty{2}{\micro\watt}, preventing potential photodegradation of the sample.

\begin{table}[h!]
\centering
\caption{Summary of Measured Peaks for Orthodontics Aligners (PET-G). \textbf{Note:} Although the overall spectral profile differs from that in reference~\cite{loskot2023influence}, several characteristic peaks—such as those near 1100~cm$^{-1}$, 1260~cm$^{-1}$, and 1719~cm$^{-1}$—are clearly present in both datasets. These features are associated with key vibrational modes of PET-G and support the spectral identification, though we acknowledge that full spectral matching would require controlled acquisition parameters and material characterization across multiple batches.}
\begin{tabular}{ccc}
\toprule
\textbf{Peak Center (cm$^{-1}$)} & \textbf{Amplitude} \\
\midrule
$\approx$ 1100 & 375\\
$\approx$ 1260 & 1152\\
$\approx$ 1600 & 9638\\
1719 &  641\\

\bottomrule
\end{tabular}
\label{tab:summary_aligners_peaks}
\end{table}

Our preprocessing workflow successfully addressed the noise challenges in the raw data, producing a clean and interpretable spectrum. The extracted peaks align with known PET-G molecular features \cite{nicita2023chemical, loskot2023influence} (see Table \ref{tab:summary_aligners_peaks}), confirming the effectiveness of our method in analyzing complex polymeric materials.

\section{Discussion}

The proposed framework offers an automated and systematic solution to many issues associated with Raman spectroscopy data analysis. However, balancing performance and automation remains challenging. Automation decreases human error risk and also increases reproducibility but at the price of increased computational complexity. Adaptive algorithms, though promising, call for extra processing power and often depend on high quality training data and well-tuned parameters.

In practical use cases, the challenge of handling fluorescence-heavy samples and noisy environments still remains. Even with baseline correction algorithms like ALS, strong fluorescence can conceal important spectral features. Additionally, high noise might prevent precise peak detection, particularly when the peaks are weak or overlapped. Consequently, future efforts must be centered on baseline correction methods and noise-resistant algorithms.

The presented framework has several advantages over conventional approaches. More specifically, it eliminates the need for manual intervention by automating crucial steps like despiking and baseline correction. Unlike polynomial fitting which calls for manual changes, ALS offers a more robust solution for dynamic fluorescence backgrounds. Voigt peak fitting is integrated to offer more accurate peak characterization with instrumental and natural broadening taken into account.

In complex Raman spectra with overlapping peaks, such as Fig.\~\textbackslash{}ref\{fig:raman\_med\}, uniqueness of the fitted solution is addressed through a combination of strategies. First, the algorithm imposes physically motivated constraints on Voigt peak shapes, widths, and amplitudes. Second, an iterative residual analysis is used to detect missing peaks after each round of fitting, and model simplification is performed by removing low-amplitude peaks that fall below a signal threshold. These procedures reduce the likelihood of overfitting and help suppress artifacts.

It is important to note that our framework is designed to be simple and modular. Rather than enforcing a fixed model, it allows straightforward modification of all key parameters, making it easy to adapt to different sample types, instruments, or analysis goals. This design choice ensures transparency and flexibility in refining peak-fitting strategies.

Regarding baseline correction, the algorithm assumes an idealized preprocessing step that includes complete suppression of noise, cosmic rays, and baseline drift. In practice, small residuals may remain, which can occasionally lead to the detection of false low-amplitude peaks. To mitigate this, the algorithm filters out peaks below a defined amplitude-to-baseline threshold. While this does not eliminate all risk of false positives, it preserves robustness for high-SNR peaks and facilitates reliable interpretation in most practical scenarios.

An important consideration in automated preprocessing workflows is parameter sensitivity—especially for algorithms such as IARPLS. While IARPLS offers improved adaptability over conventional ALS, its performance heavily depends on optimal selection of parameters such as \( \lambda \) and \( p \). To address this, recent research has emphasized automatic parameter tuning frameworks that reduce manual tuning while improving accuracy. These include generate-evaluate strategies for metaheuristics \cite{huang2020survey}, innovative cross-validation metrics based on predictive correlation \cite{holter2023r2}, and algorithm tuning frameworks structured around a problem-algorithm-tuner hierarchy \cite{eiben2011parameter}. The uniform design (UD) approach also provides an efficient alternative for parameter search, reducing computational load while maintaining accuracy \cite{wang2020uniform}. Incorporating these automated or data-driven tuning methods into IARPLS implementations would further strengthen the robustness and consistency of Raman data analysis pipelines, particularly for large-scale or high-throughput applications.

Another strength of the proposed workflow is its reproducibility. By making the code open source we encourage collaboration and transparency in the field. However, ensuring reproducibility across various datasets calls for detailed documentation and validation of the algorithms. It is acknowledged that various instruments and experimental setups will produce different results and additional standardization is needed.

The potential applications of this framework go beyond materials science to include real time diagnostics, quality control in pharmaceuticals and environmental monitoring. Automated Raman analysis could be crucial in clinical settings where quick and precise molecular characterization is needed. In the pharmaceutical industry, the ability to conduct in-line quality assessments can improve production efficiency and product consistency.

Machine learning is likely to become more essential for Raman data analysis. Particularly, deep learning models could be used to automate parameter optimization, peak detection, and classification tasks. However, large labeled datasets are still required for their practical implementation. Furthermore, the real-time applications will require lightweight algorithms that could work well under hardware constraints.

Looking ahead, future advancements in quantum-enhanced Raman spectroscopy \cite{casacio2021quantum} and hybrid preprocessing algorithms hold potential to further improve the accuracy and speed of analysis. By combining multiple techniques, such as wavelet transforms with machine learning models, it may be possible to create more resilient workflows that adapt to diverse datasets and conditions.

The existing framework has strengths but limitations. Some parameter tuning might still be necessary for very complex datasets and the iterative fitting strategy could prove computationally demanding for spectra with many overlapping peaks. Future work will focus on computational efficiency optimization of the workflow and more advanced machine learning models for real time analysis. Overall, the proposed framework offers a solid foundation for further improvement of Raman spectral analysis with wide scope for further development and application in various scientific and industrial contexts. Our contribution, which consists of open-source code made available on GitHub \cite{git}, is a step toward meeting this need, and we anticipate that the broader scientific community will continue to build upon these foundations to further advance the field.

A core strength of this work is the availability of modular Python code that mirrors the analytical workflow described throughout the manuscript. This empowers researchers to test, validate, and extend the algorithms using their own datasets or custom requirements. The framework is designed in a way that facilitates algorithmic substitutions and parameter exploration with minimal modification. For instance, users can easily replace baseline correction methods or denoising strategies by importing alternative modules.

Moreover, every major processing step—cosmic ray suppression, smoothing, peak detection, Voigt fitting—is individually callable, tested, and configurable, allowing reproducibility and rapid prototyping. This modular approach aims to lower the barrier for deploying Raman analysis pipelines in both research and industrial environments, bridging the gap between theoretical development and practical application.

It is worth emphasizing that the strength of this work lies not in introducing new algorithmic innovations, but in assembling a reproducible and user-friendly framework based on established principles. As such, we intentionally prioritized transparency, clarity, and modularity over exhaustive benchmarking. Each algorithm in the pipeline—whether for baseline correction, noise suppression, or peak fitting—has been previously validated in the literature. Therefore, we chose to focus our validation efforts on demonstrating general performance across synthetic and experimental data, leaving more exhaustive quantitative comparisons as a natural next step for specialized follow-up studies. This design philosophy aims to lower the entry barrier for newcomers in the field and foster collaborative refinement by the broader research community.

\section{Conclusion}

In this study, we established a robust and automated framework for the preprocessing and analysis of Raman spectroscopy data, focusing on baseline correction, noise suppression, and peak fitting. A key strength of this workflow lies in its simplicity and reproducibility: it is easy to apply across diverse datasets, requires minimal manual intervention, and maintains consistent performance under varied conditions.

The framework is designed as a general-purpose, modular starting point for Raman spectral analysis. It is transparent in its assumptions, avoids opaque heuristics, and can be easily adapted or extended for specialized analytical needs, including integration with machine learning or high-throughput automation. By using well-established mathematical methods such as Savitzky-Golay filtering and Voigt peak fitting, the workflow delivers accurate and interpretable results while remaining computationally lightweight.

We validated the proposed methods on synthetic and experimental Raman spectra and showed that they can reduce artifacts and improve spectral fidelity. The framework yields a high SNR ratio, allowing a better peak detection and parameter estimation needed for the precise interpretation of Raman spectra in various scientific and industrial uses.

Our work emphasizes the importance of selecting the appropriate parameters for preprocessing steps, since the improper configurations can skew spectral features and prevent correct analysis. When optimized, the proposed algorithms offer a scalable solution to increase the reliability of Raman spectral data. In addition, we offer open-source code on a public GitHub repository to encourage cooperation and additional advancement in the scientific community to allow adoption and reproducibility.

Future work will concentrate on computational efficiency of the algorithms in high-throughput contexts. This includes improving peak-fitting procedures, adding machine learning techniques for automated parameter optimization, and extending the framework to handle more complex spectral data (such as overlapping peaks or significant background interference). All these advancements might help accelerate Raman spectroscopy applications in biomedical diagnostics, pharmaceutical development, and materials science.

\section{Acknowledgments}
This work was partially supported by the following grants:\\
AFOSR: FA9550-20-1-0366, FA9550-20-1-0367, FA9550-23-1-0599\\
NIH: 1R01GM127696, 1R21GM142107, 1R21CA269099\\
NASA/FDA: 80ARC023CA002E
\section{Conflicts of Interest}
The authors declare no conflicts of interest.
\section{Data Availability Statement}
The data that support the findings of this study are available from the corresponding author upon reasonable request.

\bibliography{references}

\end{document}